\documentclass[final,5p,times,twocolumn]{elsarticle}
\usepackage{amsmath}

\usepackage{amsmath,amsfonts,amssymb}
\usepackage{graphicx} 
\graphicspath{{./images/}}
\usepackage{subcaption}  
\usepackage{epstopdf}
\usepackage{bm}
\usepackage{lineno}
\usepackage{multirow,booktabs}
\usepackage[table,xcdraw]{xcolor}

\biboptions{numbers,sort&compress}
\usepackage{xr-hyper}
\usepackage{hyperref}
\hypersetup{
    colorlinks=true,
    linkcolor=Maroon,
    citecolor=Maroon,
    urlcolor=Maroon,
    breaklinks=true,
}
\usepackage{textcomp,gensymb}
\usepackage{float}
\usepackage{listings}
\usepackage{color}
\definecolor{dkgreen}{rgb}{0,0.6,0}
\definecolor{gray}{rgb}{0.5,0.5,0.5}
\definecolor{mauve}{rgb}{0.58,0,0.82}
\usepackage{placeins} 
\usepackage{booktabs}
\usepackage{hyperref}
\usepackage{subcaption}
\usepackage{adjustbox} 

\usepackage{svg} 
\usepackage{tikz}
\usetikzlibrary{arrows.meta}
\usetikzlibrary{arrows.meta,positioning,shapes.geometric}

\makeatletter
\def\ps@pprintTitle{%
    \let\@oddhead\@empty
    \let\@evenhead\@empty
    \let\@oddfoot\@empty
    \let\@evenfoot\@empty}
\makeatother

\begin{document}

\begin{frontmatter}

\title{Good Enough is Better: Feasibility vs. Pareto-Optimality in Alloy Design}

\author[1]{Cayden Maguire}
\author[1]{Christofer Hardcastle}
\author[1]{Trevor Hastings}
\author[1,2,3]{Raymundo Arr\'oyave}
\author[1]{Brent Vela \corref{cor1}}

\address[1]{Department of Materials Science and Engineering, Texas A\&M University, College Station, TX 77843, USA}
\address[2]{J. Mike Walker '66 Department of Mechanical Engineering, Texas A\&M University, College Station, TX 77843, USA}
\address[3]{Wm Michael Barnes '64 Department of Industrial and Systems Engineering, Texas A\&M University, College Station, TX 77843, USA}

\cortext[cor1]{Corresponding author email: \texttt{brentvela@tamu.edu}}

\begin{abstract}
In alloy design, the search for candidate materials is often framed as an optimization problem, with the goal of identifying Pareto-optimal solutions across multiple objectives. However, Pareto-optimal solutions do not necessarily satisfy all minimum performance thresholds required for practical deployment. An alternative approach is to treat alloy design as a constraint satisfaction problem, in which the goal is to identify any solution that meets all bare minimum requirements across multiple quantities of interest. These approaches have yet to be benchmarked against each other in the context of realistic alloy design problems. In this work, we demonstrate that, in realistic alloy design campaigns involving multiple objectives and constraints, the constraint satisfaction framework yields a higher likelihood of finding viable alloys than optimization-based approaches. Furthermore, constraint-satisfaction approaches find the first viable alloy solutions earlier than optimization. Our results suggest that focusing on feasibility rather than optimality can lead to more actionable outcomes in materials discovery, particularly in highly constrained applications.
\end{abstract}


\end{frontmatter}

\section{Introduction}
For applications in aerospace, energy, and other safety-critical domains, new alloys must undergo rigorous qualification and certification before deployment \cite{khatamsaz2023bayesian}. Traditionally, alloy discovery efforts have been framed as optimization problems, seeking to identify Pareto-optimal solutions that balance multiple competing objectives \cite{khatamsaz2022multi}. However, in many real-world scenarios, the practical value of such “optimal” candidates is limited if they fail to satisfy strict minimum performance thresholds imposed by application requirements \cite{osti_2305503}. This raises a fundamental question: Does alloy design need optimality, or should the focus shift toward identifying any alloy that is simply good enough to meet all necessary constraints?

In the materials design literature, multi-objective optimization has been the dominant paradigm, enabled by advances in Bayesian optimization, evolutionary algorithms, and other surrogate-modeling techniques \cite{khatamsaz2022multi,khatamsaz2023bayesian,Biswas_2021,MANNODIKANAKKITHODI201692,Myung2025}. However, the emphasis on optimal trade-offs often overlooks the reality that many candidate materials, while Pareto-efficient, are unsuitable for deployment due to failing one or more hard constraints (e.g., corrosion resistance, manufacturability, phase stability, cost, etc.) \cite{osti_2305503}. The constraints may be binary \cite{hardcastle2025physics} or thresholds \cite{ghoreishi2019multi} on continuous properties. 

The alternative design strategy is constraint satisfaction frameworks \cite{arroyave2016inverse,galvan2017constraint,abu2018efficient, vela2023high, hardcastle2025physics, sohst2022surrogate} which explicitly seek any solution that satisfies all predefined requirements, regardless of whether it is optimal in the multi-objective sense. While this perspective aligns more directly with the decision-making processes of industrial alloy qualification, its systematic evaluation within the context of high-dimensional, uncertainty-aware Bayesian materials design has been largely absent, with few exceptions \cite{hickman2022known,hickman2025anubis,ghoreishi2019multi}.

For example, in our prior work \cite{khatamsaz2023bayesian}, we deployed a feasibility-first active-learning pipeline: we devoted the early budget to learning the constraint surfaces until the Shannon entropy of the feasibility map plateaued, indicating a stable decision boundary. We then pruned candidates with low probability of satisfying all constraints and performed multi-objective optimization within the resulting feasible subspace to trace the relevant trade-offs. This two-stage strategy avoids spending effort in infeasible regions and yields Pareto sets that are directly actionable for qualification. Despite the efficiency gain with respect to a brute-force querying of constraints, there were inefficiencies in this method. There is no need to accurately map the decision boundary in regions of the design space where objective values are poor i.e. there is no need to finely resolve the boundary in regions with poor objective values, but to rapidly identify alloys that satisfy all constraints.

This issue was recently addressed with the ANUBIS method \cite{hickman2025anubis}, which uses a modified optimization acquisition function to account for the probability of feasibility (PoF), e.g. $EI \times PoF$ where EI is expected improvement. ANUBIS adopts a feasibility-aware BO paradigm, learning the unknown constraints on-the-fly with a variational Gaussian process classifier while simultaneously learning the objectives. This joint treatment focuses effort where high performance and plausible feasibility coincide, learning just enough of the boundary to discriminate promising candidates rather than mapping it everywhere.

However, this paradigm still seeks optimal (albeit feasible) points. Furthermore, all examples provided were single constraints; in practice, alloy design imposes multiple, often coupled constraints. In this paper, we raise the question: \emph{"in realistic alloy design scenarios, are we truly interested in finding the feasible Pareto set?"} We argue that, in closed-loop alloy design, the practical goal is neither to train the most accurate regressor nor to chase a notional “utopia” composition, but to minimize the time it takes to find the \emph{first} feasible alloy to identify a material that satisfies all non-negotiable requirements rapidly. We substantiate this perspective with a benchmark comparing Bayesian multi-constraint satisfaction against Bayesian multi-objective (constrained) optimization. Additionally, we show that the time to find the first feasible alloy can be reduced by equipping GPCs with informative prior mean functions.

Constraint satisfaction naturally aligns with stakeholder priorities: certify that a candidate quickly meets pre-registered, application-specific minimum thresholds. Faster feasibility discovery reduces program risk and preserves the runway for scale-up, qualification, and integration \cite{ARPAE_StrategicVision_2022,SmolnikBergmann2020_StageGate,Ashby2011}. Moreover, in black-box, high-dimensional, discrete spaces, global Pareto optimality is not practically provable; at best, one can certify local non-dominance within the sampled set \cite{Ehrgott2005_MulticriteriaOptimization}, or one can run statistical tests for "Bayesian advantage" by comparing principled optimization outcomes against a baseline random experiment selection policy~\cite{hastings2024accelerated}. By contrast, feasibility is auditable: a candidate either satisfies the thresholds (under uncertainty-aware acceptance tests) or does not. This reframes success around certifiable outcomes, aligns with decision-maker needs, and provides a natural, decision-relevant stopping rule for alloy discovery.

\section{Methods}

\subsection{Gaussian Process}\label{sec:error_metrics} 

Gaussian processes (GPs) are common surrogates for Bayesian active learning and Bayesian constraint satisfaction. A surrogate model in this context is trained on a limited set of expensive ground-truth evaluations and replaces the expensive black-box experiments or simulations by providing a calibrated posterior over functions over the design space $\mathcal{X}$. For any design $\mathbf{x}\in\mathcal{X}$, the GP predictive distribution is Gaussian with mean $\mu(\mathbf{x})$ and standard deviation $\sigma(\mathbf{x})$, quantifying both our best estimate and its uncertainty. These uncertainties enable principled exploration, risk-aware selection, and efficient allocation of experimental and computational budgets to optimize objectives and learn constraints. In other words, the output of a GP is a normal distrubtion at any point $\mathbf{x}\in\mathcal{X}$, and based on all the normal distributions within $\mathcal{X}$ principled exploration and design of experiments can be done.


The posterior predictive distribution (i.e. the output of a trained GP) is defined by the equations below. An untrained GP is specified by a prior mean function $m(\mathbf{x})$ and a positive-definite covariance (kernel) $k(\mathbf{x},\mathbf{x}')$. Let $\mathbf{X}=\{\mathbf{x}_i\}_{i=1}^N\subset\mathcal{X}$ denote the training designs with noisy observations $\mathbf{y}\in\mathbb{R}^N$ (i.i.d.\ Gaussian noise with variance $\sigma_n^2$). For any test design $\mathbf{x}^*\in\mathcal{X}$, the posterior is Gaussian with mean and variance
\begin{equation}
\mu(\mathbf{x}^*) \;=\; m(\mathbf{x}^*) \;+\; \mathbf{k}_*^{\!\top}\!\left[\mathbf{K}+\sigma_n^2\mathbf{I}\right]^{-1}\!\big(\mathbf{y}-m(\mathbf{X})\big),
\end{equation}
\begin{equation}
\sigma^2(\mathbf{x}^*) \;=\; k(\mathbf{x}^*,\mathbf{x}^*) \;-\; \mathbf{k}_*^{\!\top}\!\left[\mathbf{K}+\sigma_n^2\mathbf{I}\right]^{-1}\!\mathbf{k}_*,
\end{equation}
where $\mathbf{K}_{ij}=k(\mathbf{x}_i,\mathbf{x}_j)$ and $\mathbf{k}_*=[\,k(\mathbf{x}_1,\mathbf{x}^*),\ldots,k(\mathbf{x}_N,\mathbf{x}^*)\,]^\top$. A nonzero prior mean can be handled directly via $m(\cdot)$ or equivalently by fitting a zero-mean GP to residuals $\mathbf{y}-m(\mathbf{X})$; both yield the above expressions. In this work, kernel hyperparameters are estimated by maximizing the log marginal likelihood.

\subsection{Gaussian Process Kernels}\label{sec:kernels}

We model covariance with the radial basis function (RBF, squared-exponential) kernel due to its smoothness and robustness for nonlinear response surfaces over composition. In its ARD (automatic relevance determination) form,
\begin{equation}
K_{\mathrm{RBF}}(x,x') \;=\; \sigma_f^2 \exp\!\Bigg(-\tfrac{1}{2}\sum_{d=1}^{D} \frac{(x_d - x_d')^2}{\ell_d^{\,2}}\Bigg),
\end{equation}
where $\sigma_f^2$ is the signal variance and $\ell_d$ is a per-feature length-scale.

We use ARD to assign an independent length-scale to each elemental fraction (feature). Compositions are expressed in at.\%; accordingly, we initialize $\ell_d$ in composition units and constrain them to broad bounds (e.g., $30$–$100$ at.\%), letting marginal-likelihood optimization tune each $\ell_d$ to the data and grid resolution.

To capture observation noise, we add an independent white-noise term:
\begin{equation}
k_{\text{total}}(x,x') \;=\; k_{\mathrm{RBF}}(x,x') \;+\; \sigma_n^2\,\delta_{x x'},
\end{equation}
where $\sigma_n^2$ is the noise variance and $\delta_{x x'}$ is the Kronecker delta.

\subsection{Gaussian Process Classifier (Categorical)}\label{sec:error_metrics}

For categorical constraints, feasibility probabilities are obtained using conventional Gaussian Process Classification (GPC), referred to as categorical GPCs in this work. In this setting, we assume that for any design $\mathbf{x} \in \mathcal{X}$, the probability that the constraint is satisfied is
\begin{equation}
p(y = 1 \mid \mathbf{x}) = \sigma\!\big(f(\mathbf{x})\big),
\end{equation}
where $y \in \{0, 1\}$ is a binary feasibility indicator, with $y = 1$ corresponding to the constraint being satisfied. The latent function $f(\mathbf{x})$ is modeled as a Gaussian Process,
\begin{equation}
f(\mathbf{x}) \sim \mathcal{GP}\!\big(m(\mathbf{x}), k(\mathbf{x}, \mathbf{x}')\big),
\end{equation}
The link function $\sigma(\cdot)$ maps the GP output to probabilities i.e. the link function bounds the latent function between 0 and 1 so it can be interpreted as a probability. Here we use the logistic sigmoid,
\begin{equation}
\sigma(f) = \frac{1}{1 + e^{-f}}.
\end{equation}

\begin{figure}[htb!]
  \centering
  \includegraphics[width=\linewidth]{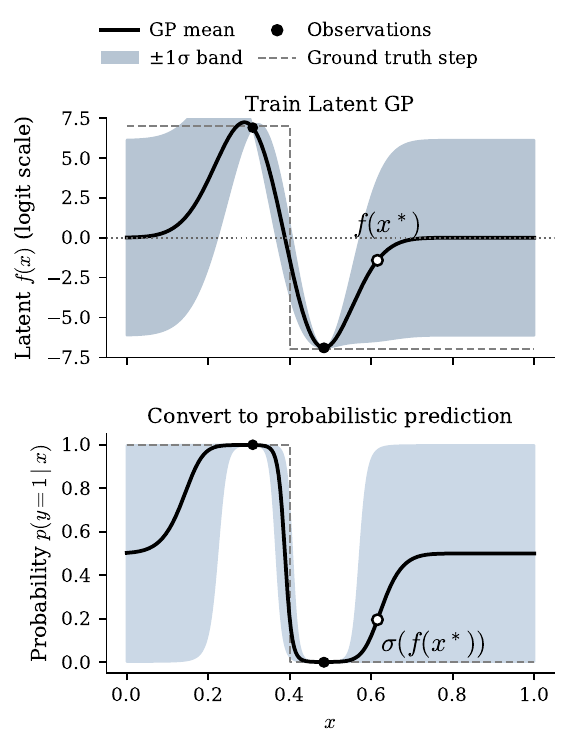}
  \caption{Categorical Gaussian Processes (GPCs) are used to model \emph{binary feasibility constraints} during both constraint satisfaction and constrained optimization. Latent GP \(f(x)\) fit to binary labels. Feasibility probability \(p(y{=}1\mid x)=\sigma\!\big(f(x)\big)\) with the same uncertainty propagated via \(\sigma(\mu_f \pm \sigma_f)\). This approach is appropriate for modeling single-phase stability, as demonstrated previously~\cite{hardcastle2025physics}, and is amenable to incorporating physics-based prior knowledge via informative prior mean functions to accelerate learning; see~\cite{hardcastle2025physics} for details.}
  \label{fig:gpc-combined}
\end{figure}

A one-dimensional example illustrating the latent GP and its categorical GP counterpart is shown in Figure~\ref{fig:gpc-combined}.


In this work, categorical GPs are used to model binary feasibility constraints during both constraint satisfaction and constrained optimization. This approach is appropriate for modeling single phase stability, as demonstrated in previous work~\cite{hardcastle2025physics}. Furthermore, this implementation is amenable to the incorporation of physics-based prior knowledge, further accelerating the learning process. More details on Gaussian Process classification with informative priors can be found in Ref.~\cite{hardcastle2025physics}.
 
\subsection{Gaussian Process Classifier (Continuous)}\label{sec:error_metrics}

For continuous constraints, feasibility probabilities are obtained using Gaussian Process (GP) regression on a real-valued quantity of interest. In this setting, for any design $\mathbf{x} \in \mathcal{X}$ we model the latent response $y(\mathbf{x}) \in \mathbb{R}$ with a GP and compute the probability that a thresholded constraint is satisfied. Let the constraint be $y(\mathbf{x}) \ge t$ (the case $y(\mathbf{x}) \le t$ follows analogously).

We place a GP prior on the latent function,
\begin{equation}
y(\mathbf{x}) \sim \mathcal{GP}\!\big(m(\mathbf{x}), k(\mathbf{x}, \mathbf{x}')\big),
\end{equation}
Given data, the GP yields a Gaussian posterior predictive distribution over $\mathbf{x}$ with mean $\mu(\mathbf{x})$ and variance $\sigma^2(\mathbf{x})$. The feasibility probability is then the posterior probability that the threshold is met:
\begin{equation}
p\big(\text{feasible}\mid \mathbf{x}\big)
\;=\;
\mathbb{P}\big(y(\mathbf{x}) \ge t \,\big|\, \mathbf{x}\big)
\;=\;
1 - \Phi\!\left(\frac{t - \mu(\mathbf{x})}{\sigma(\mathbf{x})}\right),
\end{equation}
Where $\Phi(\cdot)$ denotes the standard normal cumulative distribution function. For an upper-bound constraint $y(\mathbf{x}) \le t$,
\begin{equation}
p\big(\text{feasible}\mid \mathbf{x}\big)
\;=\;
\Phi\!\left(\frac{t - \mu(\mathbf{x})}{\sigma(\mathbf{x})}\right).
\end{equation}

\begin{figure*}[htb!] 
    \centering
    \includegraphics[width=1\linewidth]{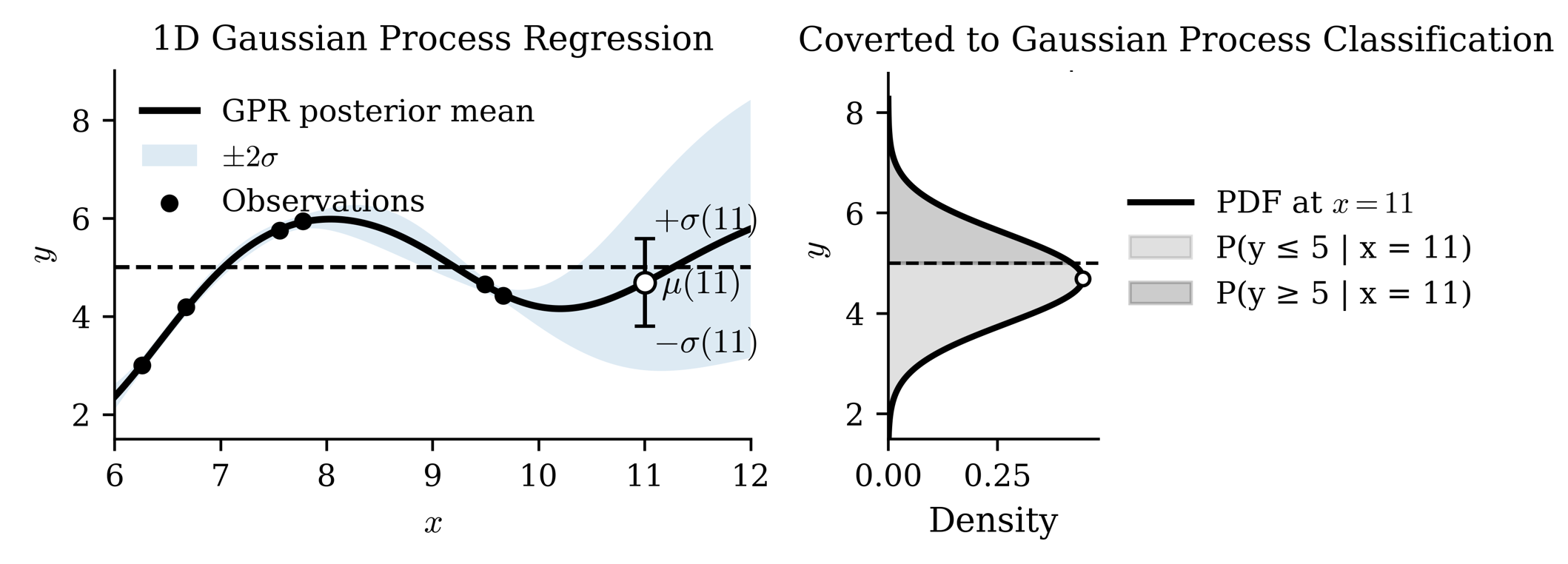}
    \caption{%
        Classification of continuous properties using a GPR. The figure on the left shows a 1D regression problem, where the black dots represent the data used to train the model and the dashed line represents an objective constraint ($y\ge5$). The mean prediction $\mu(x)$ and standard deviation $\pm \sigma(x)$ is highlighted for the arbitrary input $x=11$, and its associated posterior predictive distribution is shown on the righthand figure. Since the posterior predictive distribution is a normal distribution, the CDF can be used to dertermine the probability of meeting the constraint. 
    }
    \label{fig:contclsf}
\end{figure*}

Figure~\ref{fig:contclsf} illustrates how a Gaussian-process regressor (GPR) can be used for continuous classification via thresholding. The left panel shows observations of $y$, the posterior mean prediction \(\mu(x)\), and a credibility band (e.g., \(\pm 2\sigma(x)\)). The horizontal dashed line marks the feasibility threshold \(y=5\). The marked point at \(x=11\) highlights the model’s prediction and its uncertainty (vertical error bar). Under a Gaussian predictive marginal \(y(x)\!\sim\!\mathcal{N}(\mu(x),\sigma^2(x))\), the probability of satisfying the constraint is
\[
\mathbb{P}\big(y(11)\le 5\big) \;=\; \Phi\!\left(\frac{5-\mu(11)}{\sigma(11)}\right),
\]
which converts the continuous surrogate into a probabilistic classifier relative to the constraint.

\subsection{Probability of Satisfying All Constraints (PoF)}\label{sec:pof}
Let the design $\mathbf{x}\in\mathcal{X}$ be subject to $C$ constraints. For each constraint $c\in\{1,\dots,C\}$, let
\[
p_c(\mathbf{x}) \;=\; \mathbb{P}\big(\text{constraint $c$ is satisfied}\mid \mathbf{x}\big)
\]
denote the (calibrated) feasibility probability produced by its model: for categorical constraints via GPC,
$p_c(\mathbf{x})=\sigma\!\big(f_c(\mathbf{x})\big)$, and for continuous, thresholded constraints via GPR, 
$p_c(\mathbf{x})=\Phi\!\big((t_c-\mu_c(\mathbf{x}))/\sigma_c(\mathbf{x})\big)$ or $1-\Phi(\cdot)$ depending on the inequality direction.

Assuming conditional independence of the constraints given $\mathbf{x}$ and the learned surrogates, the overall
\emph{probability of feasibility (PoF)} is the product of the individual probabilities:
\begin{equation}
\mathrm{PoF}(\mathbf{x}) \;=\; \prod_{j=1}^{C} p_j(\mathbf{x}).
\label{eq:pof}
\end{equation}

\begin{figure}[htb!] 
    \centering
    \includegraphics[width=0.8\linewidth]{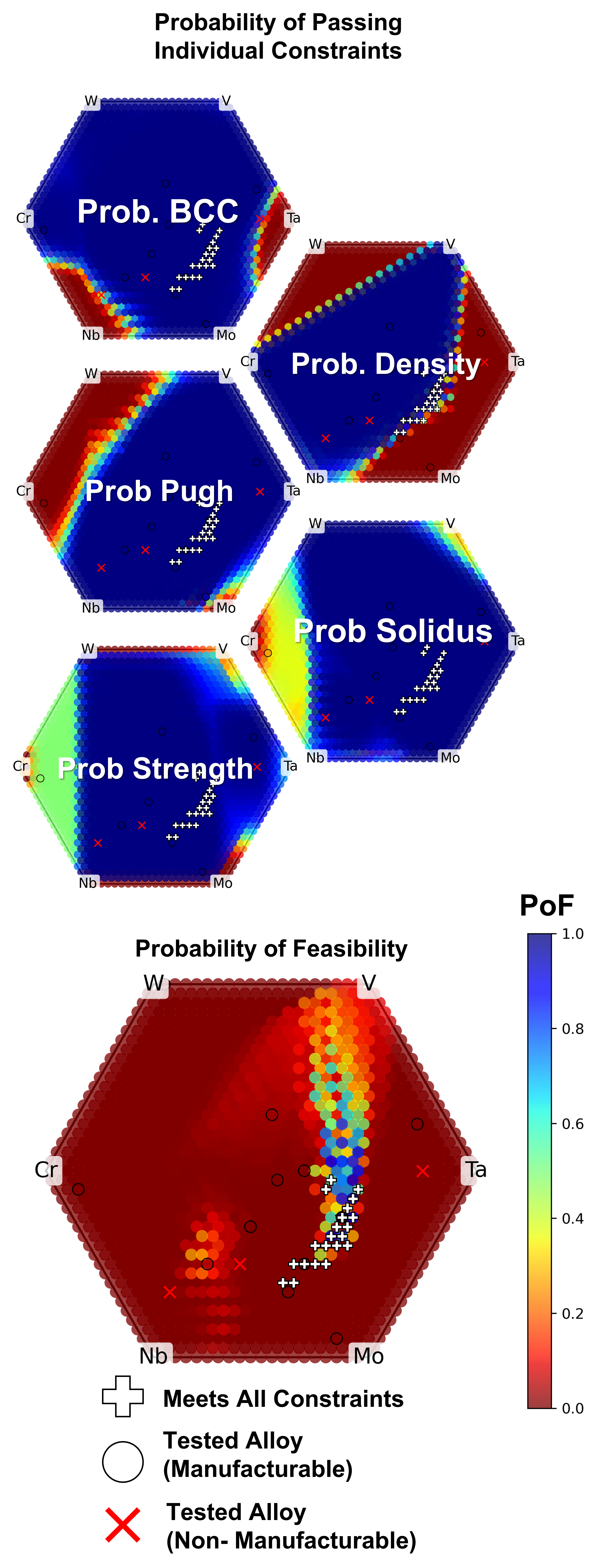}
    \caption{%
        PoF calculations across the design space. The Nb–Mo–Ta–V–W–Cr system is represented using a berycentric projection. The markers show the data used to train the GP. After training, the probability of meeting each invdividual constraint is computed across the design space (upper five projections). Assuming conditional independence, the PoF is the product of the individual probabilities (bottom projection). Note, overlap occurs on berycentric projections when datapoints are mapped closely together. This limitation is discussed in our previous work \cite{vela2025visualizing}. 
    }
    \label{fig:pof}
\end{figure}

Figure \ref{fig:pof} shows an example of PoF calculations across the design domain. The product in \eqref{eq:pof} is simple and intuitive; however, it assumes conditional independence of constraints given $\mathbf{x}$ and the fitted surrogates. When constraints are strongly correlated (e.g., share underlying physics or data), \eqref{eq:pof} can over- or underestimate the joint feasibility and become overconfident. In future work, we will replace the pure product with dependence-aware aggregators, however, the current implementation is shown to be effective for alloy design.

\subsection{Hypervolume as an Optimization Metric}\label{sec:error_metrics}

In order to benchmark constraint satisfaction against constraint optimization, meaningful metrics must be defined to compare these two campaigns. The natural metric associated with optimization is the size of the current hypervolume, which quantifies the volume of objective space dominated by the current Pareto set relative to a reference point. Hypervolume growth reflects how efficiently an optimization strategy discovers designs that improve upon previously evaluated solutions.

Formally, let $\mathbf{y}_i := \mathbf{f}(\mathbf{x}_i)$ denote the objective vector for design $\mathbf{x}_i$, and write $\mathcal{P}=\{\mathbf{y}_1,\dots,\mathbf{y}_n\}$ for the current Pareto set. Let $\mathbf{r}\in\mathbb{R}^d$ be a chosen reference point that is dominated by all feasible designs. The hypervolume indicator is defined as the $d$-dimensional Lebesgue measure (i.e., length in 1D, area in 2D, volume in 3D, and hypervolume in $d$D) of the region dominated by $\mathcal{P}$ and bounded by $\mathbf{r}$ \cite{Auger2009HypervolumeTheory}:
\begin{equation}
HV(\mathcal{P},\mathbf{r}) \;=\; \lambda_d \!\left( \bigcup_{\mathbf{y}\in \mathcal{P}} \; [\mathbf{y}, \mathbf{r}] \right),
\end{equation}
where $[\mathbf{y}, \mathbf{r}] = \{\mathbf{z}\in\mathbb{R}^d : y_i \le z_i \le r_i \ \forall i\}$ denotes the axis-aligned hyperrectangle between $\mathbf{y}$ and $\mathbf{r}$. In other words, take each Pareto point $\mathbf{y}$, draw the axis-aligned box to the reference point $\mathbf{r}$, take the union of all such boxes, and measure its size—area in 2D, volume in 3D, and hypervolume in $d$D.

\begin{figure}[t]
\centering
\begin{tikzpicture}[scale=0.8, line cap=round, line join=round]
  \draw[->] (0,0) -- (7,0) node[below] {$f_1$ (minimize)};
  \draw[->] (0,0) -- (0,6) node[left] {$f_2$ (minimize)};

  \coordinate (r) at (6,5);
  \draw[dashed] (r |- 0,0) -- (r) -- (r -| 0,0);
  \filldraw[black] (r) circle (1.5pt) node[above right] {$\mathbf r$};

  \coordinate (p1) at (1.0,4.0);
  \coordinate (p2) at (2.0,3.0);
  \coordinate (p3) at (3.0,2.2);
  \coordinate (p4) at (4.0,1.6);

  \fill[blue!20]
    (r) -- 
    (6,1.6) -- (4.0,1.6) --
    (4.0,2.2) -- (3.0,2.2) --
    (3.0,3.0) -- (2.0,3.0) --
    (2.0,4.0) -- (1.0,4.0) --
    (1.0,5) -- (6,5) -- cycle;

  \draw[dotted] (p1) -- (1.0,0) node[below] {};
  \draw[dotted] (p2) -- (2.0,0) node[below] {};
  \draw[dotted] (p3) -- (3.0,0) node[below] {};
  \draw[dotted] (p4) -- (4.0,0) node[below] {};
  \draw[dotted] (0,4.0) -- (p1);
  \draw[dotted] (0,3.0) -- (p2);
  \draw[dotted] (0,2.2) -- (p3);
  \draw[dotted] (0,1.6) -- (p4);

  \foreach \P/\name in {p1/\mathbf y_1,p2/\mathbf y_2,p3/\mathbf y_3,p4/\mathbf y_4} {
    \filldraw[black] (\P) circle (1.7pt) node[above left] {$\name$};
  }

  \node[align=left, anchor=west] at (1,5.35) {Shaded area $=$ $HV(\mathcal{P},\mathbf r)$};
\end{tikzpicture}
\caption{Bi-objective hypervolume (minimization). The shaded staircase is the region dominated by the Pareto set $\mathcal{P}$ and bounded by the reference point $\mathbf r$.}
\label{fig:hv-schematic}
\end{figure}
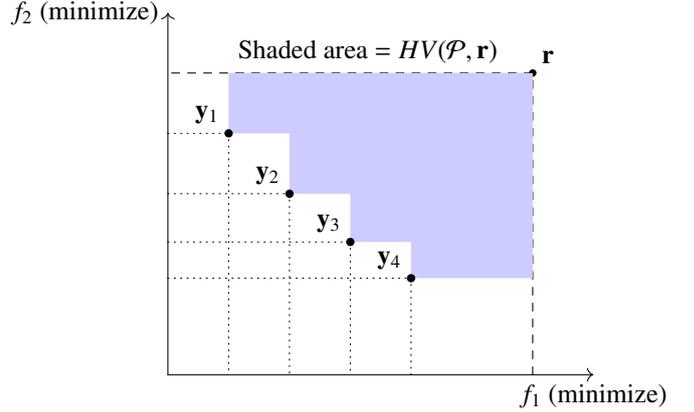

A bi-objective example of this is shown in Figure~\ref{fig:hv-schematic}. In Figure~\ref{fig:hv-schematic}, the two axes correspond to the objectives $f_1$ and $f_2$, both to be minimized. The black points $\mathbf{y}_1,\dots,\mathbf{y}_4$ represent nondominated solutions that together form the Pareto front. The dashed lines mark the reference point $\mathbf r$, which defines the bounding box for hypervolume calculation. The shaded staircase-shaped region highlights the portion of objective space dominated by the Pareto set and bounded by $\mathbf r$. The total area of this region corresponds to the hypervolume indicator $HV(\mathcal{P},\mathbf r)$, and it increases as additional nondominated points are discovered closer to the origin. This visual emphasizes how hypervolume quantifies simultaneous improvements in both objectives by capturing the volume of objective space excluded from domination by the reference point.

We will track HV as a function of iteration, wanting to maximize it as fast as possible.

\subsection{Time to First Feasible Alloy as a Constraint Satisfaction Metric}\label{sec:con_sat_error_metrics}

For constraint–satisfaction active learning, the natural progress metric is the time to discover the first feasible design. Let $C$ constraints be written as $g_c(\mathbf{x}) \le 0$ for $c=1,\dots,C$. Define the ground-truth feasibility as
\begin{equation}
F(\mathbf{x}) \;=\; \prod_{c=1}^{C} \mathbb{I}\!\big[g_c(\mathbf{x}) \le 0\big] \;\in\; \{0,1\},
\label{eq:ground-truth-feas}
\end{equation}
where $\mathbb{I}[\cdot]$ is the indicator (Iverson) function. 
This is the deterministic analogue of the \emph{probability of feasibility} $\mathrm{PoF}(\mathbf{x})=\prod_{c=1}^{C} p_c(\mathbf{x})$ introduced in Section~\ref{sec:pof}: here, the product aggregates \emph{truth values} rather than probabilities of passing. If any of the constraint $c$ is not met, the term $\mathbb{I}\!\big[g_c(x) \le 0] = 0$ and thus the entire product $F(x)=0$ for alloy $x$.


With this definition, the time required to identify the first feasible alloy (TTFF) can be expressed either in terms of iterations in the sequential case, or in the batch case, in terms of total alloys investigated. We take the perspective that the iteration-based metric is more significant, as time is the true limiting factor in practical discovery workflows. When operating in batches, the process already leverages economies of scale, reducing the marginal cost per experiment and emphasizing the importance of minimizing the number of sequential iterations rather than the total number of evaluations. However, for simplicity in this work, we benchmark sequential optimization against sequential constraint satisfaction.

\subsection{Aquisition Function for Constraint Satisfaction Campaigns}

Acquisition functions determine \emph{where} the campaign samples next, often trading off exploitation (sampling where success is most likely) and exploration (sampling to reduce uncertainty).

In our constraint–satisfaction setting, we target the feasible set of alloys with an acquisition function,
\begin{equation}
a(\mathbf{x}) \;=\; \mathrm{PoF}(\mathbf{x}),
\label{eq:acq-pof-plus-unc}
\end{equation}
where $\mathrm{PoF}(\mathbf{x})\in[0,1]$ by definition.

Since only a small percentage of alloys ($0.05\%$ of the design space) with similar properties are feasible, an exploration term is omitted from the acquisition function. 

\subsection{Aquisition Function for Optimization}

Regarding multi-objective constrained optimization, if the goal is to find Pareto optimal points and expand/improve the Pareto front, then the AQF should expect hypervolume improvement. For a candidate designs $\mathbf{x}$ with predictive distribution 
\[
\mathbf{f}(\mathbf{x}) \sim \mathcal{N}\big(\boldsymbol{\mu}(\mathbf{x}),\, \boldsymbol{\Sigma}(\mathbf{x})\big),
\]
the hypervolume improvement relative to the current Pareto set $\mathcal{P}$ and reference point $\mathbf{r}$ is
\[
\Delta \mathrm{HV}(\mathbf{y}) 
= \max\!\left\{0,\; \mathrm{HV}\!\big(\mathcal{P}\cup\{\mathbf{y}\};\mathbf{r}\big) - \mathrm{HV}(\mathcal{P};\mathbf{r})\right\}.
\]
The EHVI is the expectation of this improvement under the multivariate normal predictive distribution:
\[
\mathrm{EHVI}(\mathbf{x})
= \int_{\mathbb{R}^m} \Delta \mathrm{HV}(\mathbf{y}) \;
\mathcal{N}\!\big(\mathbf{y}; \boldsymbol{\mu}(\mathbf{x}), \boldsymbol{\Sigma}(\mathbf{x})\big)\, d\mathbf{y}.
\]

Several forms and approximations exist to calculate EHVI; however for speed (due to the computation expense), we opt to use pEHVI, achieving near-exact accuracy with markedly reduced runtime \cite{repec:spr:jglopt:v:91:y:2025:i:1:d:10.1007_s10898-024-01436-7}.

The \emph{pointwise} EHVI (pEHVI) computes, for each anchor point $\mathbf{y}^{(i)}\in\mathcal{P}$, the EHVI as if $\mathbf{y}^{(i)}$ were the only Pareto point, then takes the minimum across anchors:
\begin{equation}
\mathrm{pEHVI}(\mathbf{x}) \;=\; \min_{\,\mathbf{y}^{(i)}\in\mathcal{P}} \;\mathrm{EHVI}_{\text{1pt}}\!\big(\mathbf{x};\,\mathbf{y}^{(i)},\mathbf{r}\big),
\label{eq:pehvi-def}
\end{equation}
where $\mathrm{EHVI}_{\text{1pt}}(\cdot)$ denotes the EHVI computed with respect to the single anchored box $[\mathbf{y}^{(i)},\mathbf{r}]$ \cite{repec:spr:jglopt:v:91:y:2025:i:1:d:10.1007_s10898-024-01436-7}. Because this reduces to independent truncations along each objective, $\mathrm{EHVI}_{\text{1pt}}$ admits a closed form in terms of univariate Gaussian CDF/PDF terms, yielding a cost that is \emph{linear in $d$} for each anchor.

\newcommand{\EIi}[1]{\mathrm{EI}_i\!\left(#1\right)}
\newcommand{\zti}[1]{z_i\!\left(#1\right)}

\begin{align}
\mathrm{pEHVI}(\mathbf{x}; \mathcal{P}, \mathbf{r})
&= \min_{1 \le j \le n}\;
\mathrm{EHVI}_{\text{1pt}}\!\big(\mathbf{x};\, \mathbf{y}^{(j)}, \mathbf{r}\big), \\[4pt]
\mathrm{EHVI}_{\text{1pt}}\!\big(\mathbf{x};\, \mathbf{y}, \mathbf{r}\big)
&= \prod_{i=1}^{m} \EIi{r_i}
\;-\;
\prod_{i=1}^{m} \Big[ \EIi{r_i} - \EIi{y_i} \Big], \\
\EIi{t}
&= \big(t - \mu_i(\mathbf{x})\big)\,
\Phi\!\left( \frac{t - \mu_i(\mathbf{x})}{\sigma_i(\mathbf{x})} \right)
+ \sigma_i(\mathbf{x})\,
\phi\!\left( \frac{t - \mu_i(\mathbf{x})}{\sigma_i(\mathbf{x})} \right),
\end{align}

\subsection{Models for Ground-Truths and Priors}\label{sec:groundtruth_and_priors}

Constraint-satisfaction and multi-objective constrained-optimization campaigns used the same ground-truth models and prior models, described below.

The solidus temperature and density were calculated in Thermo-Calc using the TCHEA7 thermodynamic database using the \emph{Property} module. The prior models for these properties are rule of mixtures melting temperature and rule of mixtures density, respectively.

Phase stability was evaluated at \(600\,^\circ\mathrm{C}\) with the \textit{Equilibrium} module, selected as the target temperature for in-house hot rolling. Alloys predicted to be single-phase BCC at \(600\,^\circ\mathrm{C}\) were deemed manufacturable (a simplifying assumption used here to illustrate a simple manufacturibility constraint i.e. if this constraint is not met, other quantities of interest cannot be queried). We define “single-phase BCC” as a BCC mole fraction \(\ge 99\%\).  The prior for BCC phase stability is this work is 

\[
m_{\mathrm{BCC}}(\mathbf{x}) = 
\begin{cases}
+5, & \mathrm{VEC}(\mathbf{x}) \ge 6.87,\\
-5, & \text{otherwise},
\end{cases}
\]

which is subsequently bound between 0 and 1 with the logistic sigmoid.

We used the Maresca–Curtin model for yield strength, which posits edge-dislocation glide as the rate-controlling mechanism in BCC HEAs at elevated temperatures \cite{MARESCA2020235}. This model provides a conservative (lower-bound) estimate of yield strength for BCC RHEAs \cite{VELA2023119351} and is therefore an appropriate ground-truth model for our in~silico alloy-design exercise. The yield strength was queried at \(700\,^\circ\mathrm{C}\). The associated prior model was the Maresca-Curtin yield strength queried at \(25\,^\circ\mathrm{C}\)

As a first-order proxy for ductility, we used the Pugh ratio (\(B/G\)), computed from rule-of-mixtures estimates of the bulk and shear moduli. Prior work supports \(B/G\) as a reasonable first-approximation indicator of ductility, making it suitable for our in~silico design campaigns. We did not have a prior model for ductility in this work and instead relied on a zero-mean prior for the GP.

\subsection{Alloy Design Domain}
The design domain comprises the 6-element, 5-dimensional, Nb–Mo–Ta–V–W–Cr system. We sampled the composition simplex on a 5~at.\% grid, restricting to binaries through quinaries which yields 40{,}553 unique candidate alloys.

Within these 40{,}553 alloys, 22 (0.054\%) meet the constraints listed in Table \ref{tab:feasibility}, and 12{,}591 (31.05\%) are Pareto-optimal regarding the objectives in Table \ref{tab:optimization}. There are 18 (0.044\%) points that are both feasible and Pareto-optimal. These 22 feasible points constitute less than .1\% of the initial design space. This is consistent with prior works \cite{acemi2024multi,vela2023high} that application-relevant constraints drastically narrow the apparent breadth of the high-entropy-alloy design space


\section{Results and Discussion}

\subsection{Constraint Satisfaction Campaign}

In constraint satisfaction alloy design campaigns, the goal is to rapidly find alloys that meet all bare minimum operational constraints, i.e., minimize the TTFF. In this case study, the goal is to find alloys that meet all of the constraints in Table \ref{tab:feasibility}. This set of constraints was chosen to loosely reflect the requirements of real-world alloy design efforts \cite{ACEMI2024120379}.

\begin{table}[t]
\centering
\small
\caption{Feasibility criteria used in this work.}
\label{tab:feasibility}
\begin{tabular}{@{}ll@{}}
\toprule
Property & Criterion \\
\midrule
Solidus temperature & \(T_{\mathrm{solidus}} > 2473\,\mathrm{K}\) \;(\(2200^\circ\mathrm{C}\)) \\
Density             & \(\rho < 9.0\,\mathrm{g\,cm^{-3}}\) \\
Yield strength      & \(\sigma_{\mathrm{y}} > 700\,\mathrm{MPa}\) \\
Pugh ratio          & \(B/G > 2.5\) \\
Phase stability     & \(x_{\mathrm{BCC}} \ge 99\%\) at \(600^\circ\mathrm{C}\) \\
\bottomrule
\end{tabular}
\end{table}

To initiate the active learning scheme, first an alloy is sampled from the design space and evaluated for BCC phase stability. In our in-silico setup, we treat “processable” as single-phase BCC. If the alloy is single-phase BCC, we then measure the remaining properties in Table~\ref{tab:feasibility} and update all classifiers with the new results. If it is not single-phase BCC, we stop there and update only the BCC classifier; no other data are collected for that alloy. This mirrors real world scenarios where before investing in property tests, the alloy must be shown to be processable.

Using the trained GPCs, we estimate the probability of passing each constraint for every candidate alloy in the design space. We then multiply these probabilities to obtain the overall probability of feasibility (the chance that an alloy satisfies all constraints at once). This product assumes the constraints are conditionally independent under the model and that the classifier outputs are well-calibrated; if constraints are correlated, the result can be biased, but it remains a simple and effective screening heuristic. In future work we propose investigating the effect of accounting for these correlations.

We select the next alloy using a constraint-satisfaction aquisition function (Eqn. \ref{eq:acq-pof-plus-unc}) that favors candidates with the highest probability of feasibility. Once this point is selected for querying, the process repeats. This closed-loop is summarized in Figure \ref{fig:cs-flowchart}.

\colorlet{flowA0}{blue!0!gray}
\colorlet{flowA1}{blue!8!gray}
\colorlet{flowA2}{blue!16!gray}
\colorlet{flowA3}{blue!24!gray}
\colorlet{flowA4}{blue!32!gray}
\colorlet{flowA5}{blue!40!gray}
\colorlet{flowA6}{blue!48!gray}
\colorlet{flowA7}{blue!56!gray}
\colorlet{flowA8}{blue!64!gray}
\colorlet{flowA9}{blue!72!gray}
\colorlet{flowA10}{blue!80!gray}
\colorlet{flowA11}{blue!88!gray}
\colorlet{flowA12}{blue!96!gray}

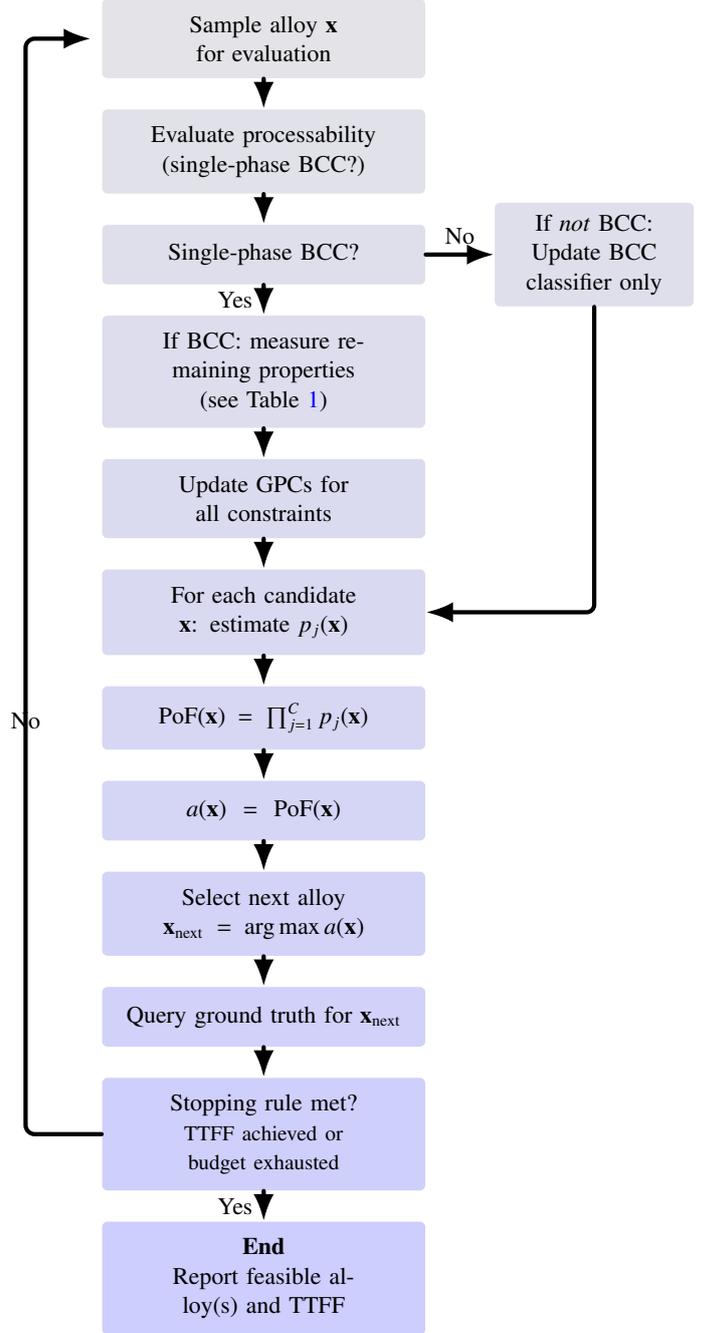
\begin{figure}[h!tb]
\centering
\begin{tikzpicture}[
  line cap=round, line join=round,
  >=Latex,
  node distance=4mm and 7mm,
  line/.style={-Latex, ultra thick},
  every node/.style={font=\small, align=center},
  block/.style={rectangle, rounded corners=2pt, draw=none,
                text width=.43\columnwidth, minimum height=5ex, inner sep=1.4ex}
]

\node[block, fill=flowA1!20!white]   (sample) {Sample alloy $\mathbf{x}$ for evaluation};
\node[block, fill=flowA2!20!white, below=of sample]
                                     (proc)   {Evaluate processability\\(single-phase BCC?)};
\node[block, fill=flowA3!20!white, below=of proc]
                                     (bcc)    {Single-phase BCC?};

\node[block, fill=flowA4!20!white, below=of bcc]
                                     (measure){If BCC: measure remaining properties\\(see Table~\ref{tab:feasibility})};
\node[block, fill=flowA5!20!white, below=of measure]
                                     (update) {Update GPCs for all constraints};
\node[block, fill=flowA6!20!white, below=of update]
                                     (pjs)    {For each candidate $\mathbf{x}$: estimate $p_j(\mathbf{x})$};
\node[block, fill=flowA7!20!white, below=of pjs]
                                     (pof)    {$\mathrm{PoF}(\mathbf{x})=\prod_{j=1}^{C} p_j(\mathbf{x})$};
\node[block, fill=flowA8!20!white, below=of pof]
                                     (acq)    {$a(\mathbf{x})=\mathrm{PoF}(\mathbf{x})$};
\node[block, fill=flowA9!20!white, below=of acq]
                                     (select) {Select next alloy $\mathbf{x}_{\text{next}}=\arg\max a(\mathbf{x})$};
\node[block, fill=flowA10!20!white, below=of select]
                                     (query)  {Query ground truth for $\mathbf{x}_{\text{next}}$};
\node[block, fill=flowA11!20!white, below=of query]
                                     (stop)   {Stopping rule met?\\\footnotesize TTFF achieved or budget exhausted};
\node[block, fill=flowA12!20!white, below=of stop]
                                     (end)    {\textbf{End}\\Report feasible alloy(s) and TTFF};

\node[block, fill=flowA3!20!white, right=9mm of bcc,
      text width=.26\columnwidth, inner sep=1ex]
  (updatebcc) {If \emph{not} BCC:\\Update BCC classifier only};

\draw[line] (sample) -- (proc);
\draw[line] (proc)   -- (bcc);

\draw[line] (bcc) -- node[left]{Yes} (measure);
\draw[line] (bcc) -- node[above]{No}  (updatebcc);
\draw[line, rounded corners=1mm] (updatebcc.south) |- (pjs.east);

\draw[line] (measure) -- (update);
\draw[line] (update)  -- (pjs);
\draw[line] (pjs)     -- (pof);
\draw[line] (pof)     -- (acq);
\draw[line] (acq)     -- (select);
\draw[line] (select)  -- (query);
\draw[line] (query)   -- (stop);

\draw[line] (stop) -- node[left]{Yes} (end);

\draw[line, rounded corners=1.2mm]
  (stop.west) -- ++(-1.cm,0)
  |- node[pos=0.18, above]{No}
  ([xshift=-1.4mm]sample.west);

\end{tikzpicture}
\caption{Closed-loop constraint–satisfaction workflow to minimize time-to-first-feasible (TTFF).}
\label{fig:cs-flowchart}
\end{figure}

\subsection{Optimization Campaign}

In multi-objective optimization, the goal is to identify and improve the Pareto front. Progress is often measured by the increase in dominated hypervolume (see Section~\ref{sec:error_metrics}). In this case study, we aim to push the Pareto front outward with respect to the objectives in Table~\ref{tab:optimization}. Similarly, these objectives were chosen to loosely mirror the needs of real-world alloy design efforts \cite{ACEMI2024120379}.

\begin{table}[t]
\centering
\small
\caption{Optimization objectives with the remaining manufacturibility constraint.}
\label{tab:optimization}
\begin{tabular}{@{}ll@{}}
\toprule
Quantity & Objective \\
\midrule
Solidus temperature & maximize \(T_{\mathrm{solidus}}\) \([\mathrm{K}]\) \\
Density             & minimize \(\rho\) \([\mathrm{g\,cm^{-3}}]\) \\
Yield strength      & maximize \(\sigma_{\mathrm{y}}\) \([\mathrm{MPa}]\) \\
Pugh ratio          & maximize \(B/G\) \\
\\
\multicolumn{2}{@{}l@{}}{\emph{Constraint:} \(x_{\mathrm{BCC}} \ge 99\%\) at \(600^\circ\mathrm{C}\).} \\
\bottomrule
\end{tabular}
\end{table}

To start the optimization loop, we again sample an alloy and apply the same BCC manufacturability gate. If it fails, we update only the phase-stability (BCC) classifier. If it passes, we observe the objective values.

With the current GPRs for objectives and the BCC GPC for the constraint, we evaluate predictions across the design space. For each alloy, we compute pointwise EHVI (pEHVI; see Section~X) and the probability of being single-phase BCC, \(p_{\mathrm{BCC}}\). Our acquisition is the product \( \mathrm{AQF}(\mathbf{x}) = p_{\mathrm{BCC}}(\mathbf{x}) \times \mathrm{pEHVI}(\mathbf{x}) \), which prioritizes candidates that both expand the Pareto front and are likely manufacturable. We pick the alloy that maximizes this quantity and iterate in a closed loop. This process is summarized in Figure \ref{fig:opt-flowchart}.

\colorlet{flowA0}{blue!0!gray}
\colorlet{flowA1}{blue!8!gray}
\colorlet{flowA2}{blue!16!gray}
\colorlet{flowA3}{blue!24!gray}
\colorlet{flowA4}{blue!32!gray}
\colorlet{flowA5}{blue!40!gray}
\colorlet{flowA6}{blue!48!gray}
\colorlet{flowA7}{blue!56!gray}
\colorlet{flowA8}{blue!64!gray}
\colorlet{flowA9}{blue!72!gray}
\colorlet{flowA10}{blue!80!gray}
\colorlet{flowA11}{blue!88!gray}
\colorlet{flowA12}{blue!96!gray}

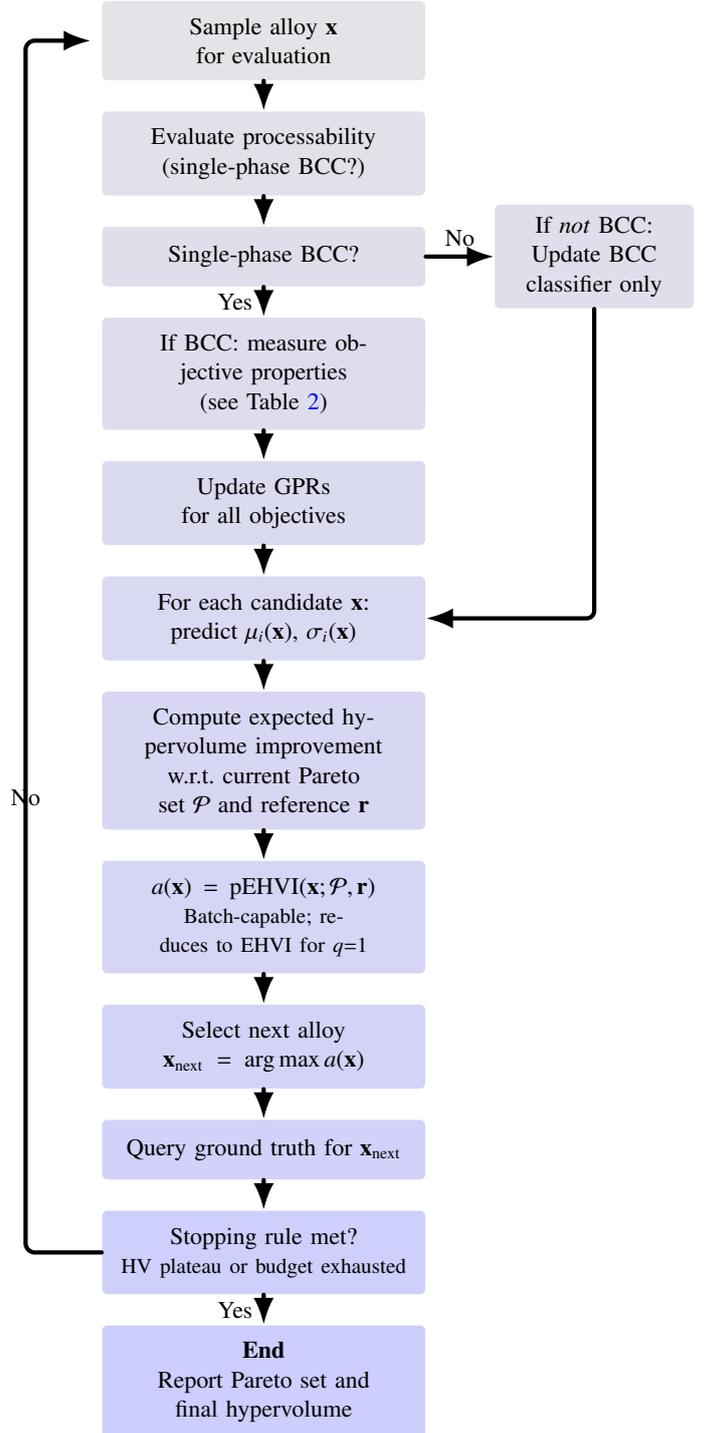
\begin{figure}[h!tb]
\centering
\begin{tikzpicture}[
  line cap=round, line join=round,
  >=Latex,
  node distance=4mm and 7mm,
  line/.style={-Latex, ultra thick},
  every node/.style={font=\small, align=center},
  block/.style={rectangle, rounded corners=2pt, draw=none,
                text width=.43\columnwidth, minimum height=5ex, inner sep=1.4ex}
]

\node[block, fill=flowA1!20!white]   (sample) {Sample alloy $\mathbf{x}$ for evaluation};
\node[block, fill=flowA2!20!white, below=of sample]
                                     (proc)   {Evaluate processability\\(single-phase BCC?)};
\node[block, fill=flowA3!20!white, below=of proc]
                                     (bcc)    {Single-phase BCC?};

\node[block, fill=flowA4!20!white, below=of bcc]
                                     (measure){If BCC: measure objective properties\\(see Table~\ref{tab:optimization})};
\node[block, fill=flowA5!20!white, below=of measure]
                                     (update) {Update GPRs for all objectives};
\node[block, fill=flowA6!20!white, below=of update]
                                     (pjs)    {For each candidate $\mathbf{x}$: predict $\mu_i(\mathbf{x}),\,\sigma_i(\mathbf{x})$};
\node[block, fill=flowA7!20!white, below=of pjs]
                                     (ehvi)   {Compute expected hypervolume improvement\\w.r.t.\ current Pareto set $\mathcal{P}$ and reference $\mathbf{r}$};
\node[block, fill=flowA8!20!white, below=of ehvi]
                                     (acq)    {$a(\mathbf{x})=\mathrm{pEHVI}(\mathbf{x};\mathcal{P},\mathbf{r})$\\
                                                \footnotesize Batch-capable; reduces to EHVI for $q{=}1$};
\node[block, fill=flowA9!20!white, below=of acq]
                                     (select) {Select next alloy $\mathbf{x}_{\text{next}}=\arg\max a(\mathbf{x})$};
\node[block, fill=flowA10!20!white, below=of select]
                                     (query)  {Query ground truth for $\mathbf{x}_{\text{next}}$};
\node[block, fill=flowA11!20!white, below=of query]
                                     (stop)   {Stopping rule met?\\\footnotesize HV plateau or budget exhausted};
\node[block, fill=flowA12!20!white, below=of stop]
                                     (end)    {\textbf{End}\\Report Pareto set and final hypervolume};

\node[block, fill=flowA3!20!white, right=9mm of bcc,
      text width=.26\columnwidth, inner sep=1ex]
  (updatebcc) {If \emph{not} BCC:\\Update BCC classifier only};

\draw[line] (sample) -- (proc);
\draw[line] (proc)   -- (bcc);

\draw[line] (bcc) -- node[left]{Yes} (measure);
\draw[line] (bcc) -- node[above]{No}  (updatebcc);
\draw[line, rounded corners=1mm] (updatebcc.south) |- (pjs.east);

\draw[line] (measure) -- (update);
\draw[line] (update)  -- (pjs);
\draw[line] (pjs)     -- (ehvi);
\draw[line] (ehvi)    -- (acq);
\draw[line] (acq)     -- (select);
\draw[line] (select)  -- (query);
\draw[line] (query)   -- (stop);

\draw[line] (stop) -- node[left]{Yes} (end);

\draw[line, rounded corners=1.2mm]
  (stop.west) -- ++(-1.cm,0)
  |- node[pos=0.18, above]{No}
  ([xshift=-1.4mm]sample.west);

\end{tikzpicture}
\caption{Closed-loop multi-objective optimization workflow using pEHVI under a single-phase BCC constraint.}
\label{fig:opt-flowchart}
\end{figure}

\subsection{Constraint Satisfaction Minimizes TTFF}
\begin{figure}[hbt!]
  \centering
  \includegraphics[width=\columnwidth]{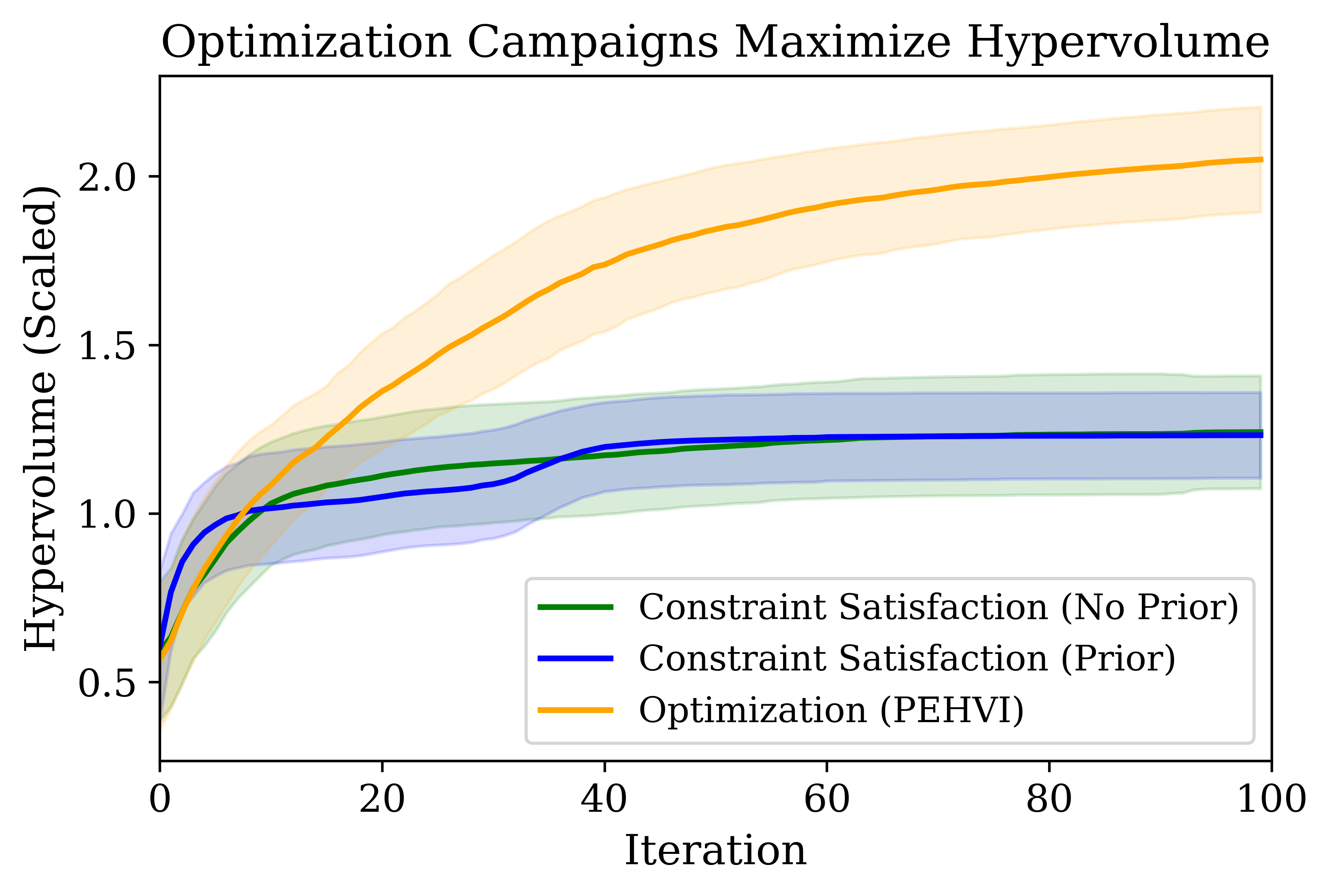}
  \caption{Average dominated hypervolume per interation. The solid lines show the average metrics for each iteration, with the optimzation shown in orange, the constraint satisfaction with a prior shown in blue, and the constraint satisfaction without a prior shown in green. The shaded region shows one standard deviation above and below the mean.}
  \label{fig:hv}
\end{figure}

\begin{figure}[hbt!]
  \centering
  \includegraphics[width=\columnwidth]{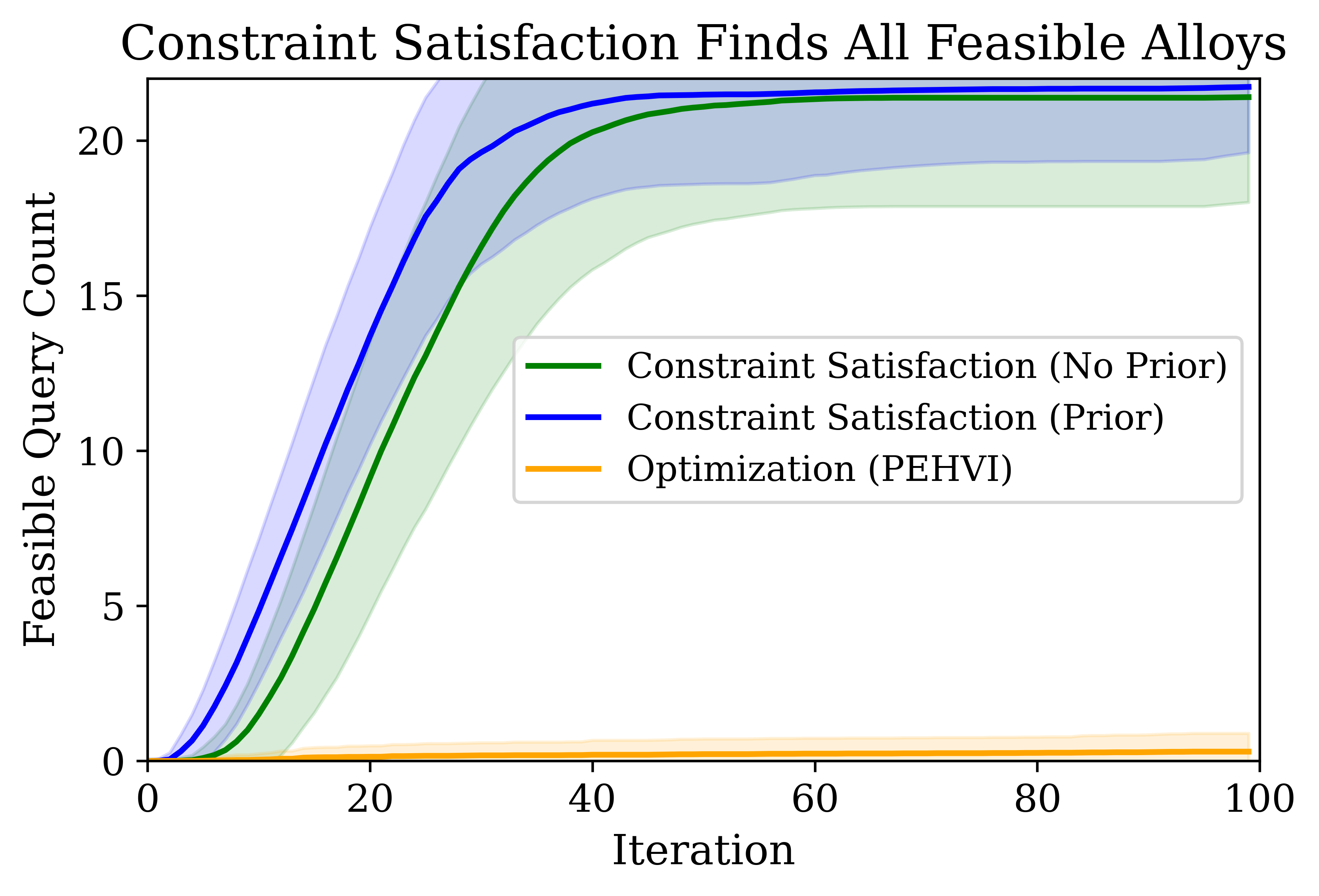}
  \caption{Average cumulative feasible alloys found per iteration. The solid line shows the average cumulative queries, with the optimzation shown in orange, the constraint satisfaction with a prior shown in blue, and the constraint satisfaction without a prior shown in green. The shaded region shows one standard deviation above and below the mean. There are 22 feasible alloys out of 41,496 possible candidates.}
  \label{fig:feasible}
\end{figure}

In order to provide a measure of statistical robustness and to make sure any effect of initialization, hyperparameter tuning, etc., is accounted for, we run 200 of these closed loops for optimization and constraint satisfaction and track the metrics listed in Sections \ref{sec:error_metrics} and \ref{sec:con_sat_error_metrics} as a function of iteration. The mean and standard deviation of these metrics vs iteration is plotted in \ref{fig:hv} and \ref{fig:feasible}.


Figure \ref{fig:hv} shows dominated hypervolume versus iteration, and Figure \ref{fig:feasible} shows the cumulative number of feasible alloys found. As expected in \ref{fig:hv}, the optimization strategy increases hypervolume more rapidly, whereas in Figure \ref{fig:feasible} constraint satisfaction discovers feasible points sooner and in greater numbers. By the end of the experiment, on average, the constraint-satisfaction loop identifies all but one of the feasible alloys and its average time-to-first-feasible (TTFF) is 8 iterations. Recall that only 22 out of 41,496 candidates are feasible (0.05\% of the design space), making this a needle-in-a-haystack problem solved within a short campaign (half of the feasible alloys found by iteration 17; 21 out of 22 feasible alloys found by iteration 35).

\subsection{Informative Prior Mean Functions Minimize TTFF}

Having shown that constraint satisfaction achieves a lower time-to-first-feasible (TTFF) than optimization, we next tested whether adding informative priors to the GPCs (see Section \ref{sec:groundtruth_and_priors}) further helps. We ran an additional constraint-satisfaction campaign in which the GPCs used informative prior mean functions, repeating the closed loop 200 times and averaging metrics by iteration. As shown in Figure \ref{fig:feasible}, the prior-equipped variant yields a smaller TTFF than the baseline (4 iterations instead of 8) and also shortens the time required to discover all feasible points. On average, the prior-equipped constraint-satisfaction loop identifies all feasible alloys by iteration 56.

\section{Discussion}

The empirical results presented here strongly suggest that, in realistic alloy design scenarios, the definition of success depends heavily on the operational context. When the discovery objective is to rapidly identify any alloy—or any other material, as these results are materials-agnostic and broadly generalizable—meeting all minimum performance thresholds, a feasibility-driven strategy yields more decision-relevant outcomes than classical multi-objective optimization. The \textit{time-to-first-feasible} (TTFF) metric introduced here captures this notion quantitatively, providing a direct measure of discovery efficiency aligned with realistic, stakeholder-driven priorities.

Constraint satisfaction reframes alloy design around \textit{certifiable feasibility} rather than \textit{unverifiable optimality}. In practical alloy development, the ground-truth material response surfaces are never fully known; consequently, global Pareto optimality cannot be proven—it can only be approximated within the sampled design space. At best, one can certify local non-dominance among observed candidates, a fact that limits the decision value of optimization-based strategies in finite-budget campaigns. By contrast, feasibility is auditable: a candidate either satisfies the registered performance thresholds (within uncertainty bounds) or it does not. \emph{The difference between feasibility (which we can test and certify) and optimality (which we can only approximate) marks the fundamental limit of what can be known in materials discovery---its epistemic boundary---when our information is incomplete or uncertain.}

The observed impact of informative prior mean functions further supports the value of embedding physical intuition into probabilistic classifiers. By incorporating thermodynamic or mechanical priors, we showed that Gaussian Process Classifiers (GPCs) can learn feasibility boundaries more efficiently and focus querying on physically meaningful regions of the design space. This approach naturally extends to multi-fidelity and cost-aware frameworks, in which prior knowledge derived from simulations or relevant prior experiments accelerates feasibility discovery.

Finally, the complementarity between constraint satisfaction and optimization suggests that these paradigms should not be viewed as competing but sequential. A hybrid strategy, beginning with feasibility discovery and transitioning to Pareto refinement, mirrors the logic of stage-gate design processes in engineering practice. In this structure, constraint satisfaction would provides the \textit{go/no-go} certainties required for decision-making, while optimization explores trade-offs within the feasible subspace. The combination of these complementary approaches constitute a unified, fully-Bayesian workflow that effectively balances exploration and exploitation over the long range within a materials discovery campaign.

\section{Conclusion}

In this work, we benchmarked two Bayesian closed-loop discovery strategies on the same alloy design space: (i) \emph{constraint satisfaction}, which targets rapid discovery of any alloy meeting all feasibility criteria, and (ii) \emph{multi-objective optimization}, which seeks to expand the Pareto front. Across 200 independent runs per setting, we tracked dominated hypervolume and feasibility metrics as a function of iteration.

\begin{itemize}
  \item \textit{Time-to-first-feasible (TTFF).} Constraint satisfaction consistently achieved a lower TTFF than optimization, identifying feasible alloys earlier and in greater numbers. By iteration 56, the prior equiped model had, on average, recovered all 22 feasible candidates (\(0.05\%\) of the \(41,496\) total), effectively resolving a needle-in-a-haystack problem within a limited number of iterations.
  \item \textit{Hypervolume growth.} As expected, optimization expanded dominated hypervolume faster, reflecting its objective.
  \item \textit{Informative priors.} Equipping the GPCs with informative prior means further reduced TTFF and shortened the time required to find all feasible points relative to the baseline constraint-satisfaction loop.
\end{itemize}

When the priority is to obtain \emph{any} viable alloy quickly (e.g., early-stage, resource-constrained campaigns), constraint satisfaction is preferable. When the goal shifts to improving trade-offs among properties, multi-objective optimization is the right tool. In practice, a hybrid schedule—using constraint satisfaction to secure the first feasible alloy, then handing off to optimization—balances speed with frontier improvement. When prior knowledge exists, informative priors are a low-cost way to accelerate feasibility discovery.

Our feasibility model assumes approximate independence among constraints and well-calibrated classifier outputs; deviations from either can lead to systematic over- or under-estimation of the probability of feasibility ($PoF$). The BCC constraint serves as a practical proxy for manufacturability \textit{in~silico}, but it may not fully capture the spectrum of processing risks encountered experimentally. Another limitation is that the scope of the work is limited to a sequential evaluation strategy; batch selection methods were not considered. These caveats motivate future work on joint (correlated) constraint models, improved probabilistic calibration, cost-aware and multi-fidelity acquisition strategies, and adaptive hybrid frameworks that transition from feasibility-first discovery to Pareto frontier expansion as evidence accumulates. More broadly, these findings delineate the epistemic limits of optimization in materials discovery: while feasibility can be demonstrated, optimality can only be inferred.

\section*{Code Availability}

The code associated with this work is available at the following repository:
\href{https://github.com/hardccw643/constraint-satisfaction.git}{https://github.com/hardccw643/constraint-satisfaction.git}.

\section*{Declaration of Generative AI and AI-assisted technologies in the writing process}

During the preparation of this work, the authors used GPT-5 in order to ideate/brainstorm alternative sentence structures and stylistic choices in limited sections of the paper. After using this tool/service, the authors reviewed and edited the content as needed and take full responsibility for the content of the publication.

\section*{Declaration of Competing Interest}
The authors declare that they have no known competing financial interests or personal relationships that could have appeared to influence the work reported in this paper.

\section*{Acknowledgements}
We acknowledge the support from the U.S. Department of Energy (DOE) ARPA-E ULTIMATE Program through Project DE-AR0001427. RA also acknowledges the Army Research Laboratory (ARL) for support through Cooperative Agreement Number W911NF-22-2-0106, as part of the High-throughput Materials Discovery for Extreme Conditions (HTMDEC) program as supported by the BIRDSHOT Center at Texas A\&M University. BV acknowledges the support of NSF through Grant No. 1746932 (GRFP) and 1545403 (NRT-D3EM). Computations were conducted at the Texas A\&M University High-Performance Research Computing (HPRC) facility.

\section*{CRediT authorship contribution statement}
\textbf{Cayden Maguire}: Writing – original draft, Writing – review \& editing, Visualization, Software, Investigation, Formal analysis, Data curation. \textbf{Christofer Hardcastle}: Writing – original draft, Writing – review \& editing, Visualization, Software, Investigation, Formal analysis, Data curation. \textbf{Trevor Hastings}: Writing – review \& editing, Supervision. \textbf{Raymundo Arróyave}: Writing – review \& editing, Project administration, Funding acquisition. \textbf{Brent Vela}: Writing – original draft, Writing – review \& editing, Visualization, Software, Investigation, Formal analysis, Data curation,
Conceptualization, Methodology, Supervision.


\bibliographystyle{elsarticle-num}
\bibliography{mybibfile}

\end{document}